\documentclass[aps,pre,showpacs,showkeys,twocolumn]{revtex4}
\usepackage{natbib}
\usepackage{epsf} 
\usepackage{amsmath,amssymb,graphicx}
\usepackage{bm}

\begin{document}

\title{Adiabatic invariants for the dynamics of rotating 3-dimensional fluid}

\author{Alexander M. Balk}

\affiliation{Department of Mathematics, University of Utah, Salt Lake City, UT 84112}

\date{\today}

\begin{abstract}
The 3-dimensional fluid dynamics under rapid rotation with beta effect is shown to possess three adiabatic-type invariants; their presence leading to the energy accumulation in zonal jets.
\end{abstract}

\keywords{Waves, Triad-resonance interactions, Adiabatic invariant, Euler's equation, 3D fluid dynamics, Zonal jets}

\pacs{47.35.-i 	
     ,  47.32.Ef 	
     ,  92.10.Ei 	
     ,  92.10.Hm 	
     ,  52.35.Kt 	
     }     

\maketitle 

\subsection{Introduction}

The present note is an addendum to the recent paper \cite{BHW}, which described three adiabatic-type invariants for the {\it rotating shallow water} dynamics; their presence implying the energy transfer towards zonal jets. The addendum is due to the remarks of L. Smith \cite{L.Smith}. She remarked that it would be interesting if the extra conservation held for 3-dimensional (3D) fluid dynamics and expressed belief that the 3D extension would be possible. However, this seemed highly improbable, because the adiabatic conservation --- related to the conservation in triad resonances --- is essentially tied to the two-dimensionality. And even if there were 3D extension of some adiabatic invariant, say enstrophy, the resulting invariant would hardly be positive definite; and so, it would be useless for the inverse cascade argument. 

Separately, application of \cite{BHW} to the real oceans and atmospheres of rotating planets encounters the following natural question. In a fluid layer of finite depth $H$, the shallow water motions [which are 2-dimensional (2D)] can generate --- due to nonlinearity --- the 3D inertia waves. One can argue that the length of a typical Rossby wave significantly exceeds the depth $H$, and so, the energy exchange between Rossby waves and inertia waves can be disregarded due to the disparity of scales. But what about short Rossby waves? Their wavelength can be comparable to the depth $H$. 
Therefore, it appears crucial whether the extra conservations can be extended to the 3D fluid dynamics. 

Surprisingly, the 3D extension has turned out to be true. It is due to the fact discovered by H.P. Greenspan long time ago \cite{Green}. In the absence of the beta effect (i.e. on the $f$-plane), he showed that a 3D fluid flow can be decomposed into a {\it single} 2D geostrophic mode and an infinite number of 3D inertia waves [see Eq. (\ref{GenSol}) below] and found that ``the inertia modes do not produce triad-resonance response on the geostrophic mode''. It is important that there is {\it only one} geostrophic mode; otherwise, the enstrophy (which is, roughly speaking, the quadratic zonal momentum) would not be positive definite, and the inverse cascade argument would fail. The third adiabatic conservation would also fail, and so would do the balance argument for the energy transfer towards zonal jets.

Consider the dynamics of 3D inviscid fluid of constant density; the fluid velocity ${\bf v }(x,y,z,t)$ satisfying the equations
\begin{subequations}\label{Euler}
\begin{align}
{\bf v}_t+({\bf v}\cdot\nabla){\bf v}+{\bf f}\times{\bf v}&                      
=-\nabla\Pi,\\ 
\nabla\cdot{\bf v}&=0.
\end{align}
\end{subequations}
Euler's equation (\ref{Euler}a) involves Coriolis force; ${\bf f}$ is the double angular velocity; $\Pi$ is the fluid pressure (centrifugal force included) divided by the density. The fluid is assumed to occupy a layer between parallel horizontal planes, rotating about vertical axis $z$. The boundary condition is vanishing of the vertical fluid velocity at the planes. 
 
The system (\ref{Euler}) is known to conserve the energy and helicity. Here we show that in the presence of the beta effect --- when ${\bf f}=(f_0+\beta_0 y){\bf \hat z}$ slowly depends on the longitudinal coordinate $y$ ($f_0$ is a reference value of the Coriolis parameter) --- there are three integrals that are conserved by the Euler dynamics (\ref{Euler}) {\it adiabatically}. [The adiabatic conservation means {\it approximate} conservation over {\it long time}. It is often due to the {\it slow time-dependence} of some parameters, but here --- like in \cite{BHW} --- it is due to the {\it small nonlinearity}, compared to the beta-effect, and the {\it slow spatial dependence} 
${\bf f}(y)$.] 

The three adiabatic invariants are described in the next Section \ref{Sect:AdiabaticInv}, and then in Section \ref{Sect:Zonal}, it is noted that their presence implies the emergence of zonal jets.

\subsection{Adiabatic invariants}
\label{Sect:AdiabaticInv}

First, assume that the fluid equations are written in dimensionless form with time scale $1/f_0$, length scale $L$ and velocity scale $f_0 L$. Then the dimensionless typical wave amplitude is the Rossby number $U_0 / L f_0$, with $U_0$ being typical dimensional velocity. Dimensionless Coriolis parameter is ${\bf f}=(1+\beta y){\bf \hat z}$
with dimensionless parameter $\beta=L\beta_0/f_0$.

The general solution of the linearized (about ${\bf v}=0$) system (\ref{Euler}) with $\beta=0$ is a superposition of  normal modes, of which there are two types \cite{Green}: (i) an infinite number of inertial oscillations (with amplitudes $a_m$ and real non-zero frequencies $\omega_m$) 
and (ii) a single geostrophic mode ${\bf U}=(U, V)$, which is two-dimensional and steady 
\begin{align}\label{GenSol}
	{\bf v}(x,y,z,t)={\bf U}(x,y)+\sum_{m=-\infty}^{\infty} a_m {\bf V}_m(x,y,z) e^{i\omega_m t}
\end{align}
[the sum is over $m=\pm1,\pm2,\ldots\;(m=0$ is excluded); 
$\omega_{-m}=-\omega_m, \, {\bf V}_m={\bf V}_{-m}^\ast,\, a_m=a_{-m}^\ast$]. 
When the nonlinearity is present (but small) and $\beta\neq0$ (but small), the amplitudes $a_m$ and the geostrophic mode ${\bf U}$ acquire slow time-dependence. And the Euler system (\ref{Euler}) can be written as a system of evolution equations for ${\bf U}$ and amplitudes $a_m$.

For the {\it nonlinear} system with $\beta=0$, ``the inertia modes do not produce triad-resonance response on the geostrophic mode'' \cite{Green}; in other words, the coupling coefficient between the geostrophic mode (which has zero frequency) and two inertia modes (with frequencies $\omega_m$ and $\omega_n$) vanishes whenever $\omega_m+\omega_n=0$. 

When $\beta\neq 0$, the geostrophic mode becomes the quasigeostrophic mode, or Rossby wave with dispersion law
\begin{align*}
	\Omega({\bf k})=-\beta \frac{p}{k^2} \quad\quad (k^2=p^2+q^2)
\end{align*}
[${\bf k}=(p,q)$ is the horizontal wave vector]. As well, the inertia wave frequencies acquire some small corrections (corresponding to the beta correction in the Coriolis parameter). Now solving for $a_m(t)$ and 
${\bf U}(x,y,t)$ by perturbations, we encounter small denominators, which are not exactly canceled by the corresponding numerators (and could produce secular terms). 

Let us estimate the effect of nonzero $\beta$ on the small denominators. 
We will consider the mid-latitudinal beta-plane, when $\beta_0/f_0\sim 1/R_0$, $R_0$ being the Earth radius.
There are two types of triad interactions between Rossby and inertia waves:
(i) two Rossby waves and one inertia wave, and (ii) two inertia waves and one Rossby wave.

The first type interactions are clearly far from being resonant, and so, do not produce small denominators. Indeed, if $H\ll R_0$, then the Rossby frequencies $\Omega\sim\beta_0 L$ are much smaller than the inertia frequencies $\omega_m=f\,k_z/\sqrt{k^2+k_z^2}$, with vertical wave number $k_z\sim m \,\pi/H \; (m= 1, 2, \ldots)$ and horizontal wave number $k\sim\pi/L$. 

The second type can be resonant and produce small denominators.
However, the parameter $\beta\neq0$ will change small denominators only {\it slightly}: the Rossby frequencies and the change in the inertia frequencies are both $\sim\beta_0 L\ll f_0$ (at mid-latitudes). Therefore, it is possible to add $O(\beta)$-corrections to the coupling coefficients to have exact cancellation of the small denominators, and then show that the compensation for these corrections remains within the accuracy of adiabatic conservation. 
So, the rapidly rotating 3D fluid dynamics adiabatically conserves the invariants of the 2D shallow water dynamics. 

The present work is somewhat parallel to the study \cite{BHW} (containing many details) of adiabatic conservation in the rotating shallow water dynamics. It also has two types of waves: Rossby waves and inertia-gravity waves; the latter also do not produce triad-resonance response on the former. For the shallow water case, it was possible to consider situation when the correction $\beta_0 L$ to the Coriolis parameter exceeds $f_0$ (which happens, e.g., near the equator). In this situation, parameter $\beta$ should be small compared to $\sqrt{gH}/L^2$ ($g$ is the gravity constant).  It is unclear at present whether such situation admits extra conservation for the rotating {\it 3D} fluid dynamics.

The adiabatic invariants are integrals
\begin{align}\label{I}
	I=\frac{1}{2}\int X_{\bf k}\, |\zeta_{\bf k}|^2  \, d{\bf k},
\end{align}
where $\zeta_{\bf k}$ is the 2D Fourier transform of the 2D vorticity $\zeta(x,y,t)=V_x-U_y$, which can be found 
by averaging the vertical component of the 3D vorticity $\nabla\times{\bf v}$ along the vertical direction
\begin{align*}
	\zeta(x,y,t)=\frac{1}{H}\int_0^H (v_x-u_y) dz.
\end{align*}
The kernel $X_{\bf k}$ is an arbitrary linear combination of the following 3 functions
\begin{subequations}\label{kernels}
\begin{align}
	X_{\bf k}&=\frac{1}{k^2},\\
	X_{\bf k}&=1,\\
	X_{\bf k}&=\frac{p^2(p^2+5 q^2)}{k^{10}}.
\end{align}
\end{subequations}
The function (\ref{kernels}a) determines the energy of the quasigeostrophic mode, its enstrophy corresponds to the function (\ref{kernels}b). The function (\ref{kernels}c) appeared in \cite{B1991} for the dynamics described by a single quasigeostrophic equation in the limit when the Rossby radius of deformation $\rho$ approaches infinity. For the shallow water system, the limit $\rho\rightarrow\infty$ leads to small denominators of two kinds; both can be canceled \cite{BHW}; and the same function (\ref{kernels}c) determines the extra invariant in the rotating shallow water dynamics.  

The adiabatic conservation of the integral (\ref{I}) means that variation $\Delta I=I(t)-I(0)$ remains small (compared to $I$) for a long time $t$ of many wave periods. Of course, the adiabatic conservation holds if $\dot I$ is sufficiently small. However, $\Delta I$ can be small even if 
$\dot I$ is large: $I(t)$ can oscillate in time, as usually the case for adiabatic conservation in the theory of dynamical systems; and this is also the case here. 

\subsection{Emergence of zonal jets}
\label{Sect:Zonal}

In rapidly rotating 3D fluids, the energy is transferred towards 2D flow. This was demonstrated in numerical simulations \cite{SmWa}; this paper describes several possible mechanisms of such transfer beyond triad resonances. So, the energy accumulates in the 2D quasigeostrophic mode.  

Any of the kernels (\ref{kernels}) is positive, and the corresponding invariant (\ref{I}) is positive definite. The invariants (\ref{I}) with kernels (\ref{kernels}ab) imply the inverse cascade of the energy of the quasigeostrophic mode. The presence of all three adiabatic invariants (\ref{kernels}abc) implies \cite{BHW} that the energy of the quasigeostrophic mode flows to large scale {\it zonal} flow (not just to large scales, but to the region of 
the ${\bf k}$-plane with polar angles $\theta=\arctan(q/p)$ close to $\pm\pi/2$). 

Thus, the rapidly rotating 3D dynamics transfers energy towards zonal jets. 

\subsection{Conclusion}
\label{Sect:Concl}

We have found three adiabatic invariants of the rapidly rotating 3D fluid dynamics (namely the energy, enstrophy, and extra invariant of the quasigeostrophic mode). Their presence together with the {\it two-dimensionalization} (see Sec.\ \ref{Sect:Zonal}) implies the transfer of energy towards zonal jets.

It would be interesting to consider adiabatic invariants for 3D fluid layer with a {\it free surface}. This situation presents a combination of the present work with \cite{BHW} and would lead to the general Rossby wave dispersion, with a {\it finite} Rossby radius of deformation $\rho$. In that general situation, the 3D fluid dynamics still seems to have the three adiabatic invariants, which would still lead to the energy transfer towards zonal jets --- except for the long wave limit, when the forcing scale (at which the energy is supplied) is much bigger than $\rho$ (the relation between $L$ and $\rho$ can be arbitrary). 
 
\begin{acknowledgments}
I am indebted to L. Smith for stimulating remarks and subsequent discussions. As well, I am grateful to the organizers of the Geophysical Fluid Dynamics summer school (Woods Hole, Massachusetts, USA) and programs at the Aspen Center for Physics (Colorado, USA).
\end{acknowledgments}

\bibliography{My}
\end{document}